\documentclass{aastex631}

\usepackage{amsmath}
\usepackage{graphicx}
\usepackage{float}

\begin{document}

\title{Asymmetries in the simulated ozone distribution on TRAPPIST-1e due to orography}

\author{Anand Bhongade}
\affiliation{School of Earth and Environment, University of Leeds, Leeds, UK}

\author{Daniel R Marsh}
\affiliation{School of Physics and Astronomy, University of Leeds, Leeds, UK}
\affiliation{School of Chemistry, University of Leeds, Leeds, UK}

\author{Felix Sainsbury-Martinez}
\affiliation{School of Physics and Astronomy, University of Leeds, Leeds, UK}

\author{Gregory Cooke}
\affiliation{School of Physics and Astronomy, University of Leeds, Leeds, UK}
\affiliation{Institute of Astronomy, University of Cambridge, UK}



\begin{abstract}

TRAPPIST-1e is a tidally locked rocky exoplanet orbiting the habitable zone of an M dwarf star. Upcoming observations are expected to reveal new rocky exoplanets and their atmospheres around M dwarf stars. To interpret these future observations we need to model the atmospheres of such exoplanets. We configured CESM2-WACCM6, a chemistry climate model, for the orbit and stellar irradiance of TRAPPIST-1e assuming an initial Earth-like atmospheric composition. Our aim is to characterize the possible ozone (O$_3$) distribution and explore how this is influenced by the atmospheric circulation shaped by orography, using the Helmholtz wind decomposition and meridional mass streamfunction. The model included Earth-like orography and the substellar point was located over the Pacific Ocean. For such a scenario, our analysis reveals a North-South asymmetry in the simulated O$_3$ distribution. The O$_3$ concentration is highest at pressures $>$ 10 hPa (below $\sim$30 km) near the South Pole. This asymmetry arises from the higher landmass fraction in the Northern Hemisphere, which causes drag in near-surface flows and leads to an asymmetric meridional overturning circulation. Catalytic species were roughly symmetrically distributed and were not found to be primary driver for the O$_3$ asymmetry. The total ozone column (TOC) density was higher for TRAPPIST-1e compared to Earth, with 8000 Dobson Units (DU) near the South Pole and 2000 DU near the North Pole. The results emphasize the sensitivity of O$_3$ to model parameters, illustrating how incorporating Earth-like orography can affect atmospheric dynamics and O$_3$ distribution. This link between surface features and atmospheric dynamics underlines the importance of how changing model parameters used to study exoplanet atmospheres can influence the interpretation of observations.

\end{abstract}

\keywords{Exoplanet atmospheres (487); Exoplanet atmospheric composition (2021); Exoplanet atmospheric dynamics (2307); Transmission spectroscopy (2133); James Webb Space Telescope (2291)}


\section{Introduction} \label{sec:intro}

Scientists have long wondered about the existence of life on other planets and this curiosity has motivated them to explore celestial bodies beyond our solar system. To date, over 5600 confirmed exoplanets have been identified by instruments like the Kepler Space Telescope, the Transiting Exoplanet Survey Satellite, and various space-based and ground-based telescopes.\footnote{\url{https://exoplanetarchive.ipac.caltech.edu/docs/counts_detail.html} - Date: 15/04/2024} Studies utilizing Kepler mission data have found that the presence of small rocky exoplanets around M dwarf stars exceeds that around Sun-like stars \citep{howard2012planet, dressing2015occurrence, mulders2015increase, gaidos2016they}, although this is likely because of the observational bias of telescopes for detecting exoplanets around smaller, cooler and dimmer stars. To date, 200 confirmed rocky exoplanets have been detected.\footnote{\url{https://exoplanets.nasa.gov/what-is-an-exoplanet/planet-types/terrestrial/} - Date: 15/04/2024}

M dwarf stars, constituting approximately 70\% of all known stars in our galaxy, have garnered scientific attention due to their abundance and compact planetary systems \citep{Bochanski_2010}. The location of the habitable zone around a star depends upon its stellar properties \citep{huang1959problem}. M dwarf stars, characterized by their low temperature and flux, have habitable zones situated closer to them. Detection methods such as transit photometry and radial velocity rely on the planet-to-star mass and size ratio, making the search for rocky exoplanets in habitable zones around M dwarf stars more feasible due to the star’s relatively small size \citep{shields2016habitability, reiners2018carmenes,gould2003sensitivity, nutzman2008design}.

 Exoplanets orbiting M dwarfs can be in such proximity to their stars that they become tidally locked, resulting in a permanent day side and night side. Tidal locking occurs due to the gravitational force exerted by the star, which distorts the planet into an elongated shape. This results in synchronous rotation, where the planet's rotational period equals its orbital period \citep{barnes2017tidal}.

In addition to {the incident energy flux from the host star and orbital configuration, a planet's location relative to the habitable zone} is significantly influenced by atmospheric composition, particularly the presence or absence of greenhouse gases. Ozone (O$_3$), acting as a greenhouse gas on Earth, influences the vertical temperature structure and humidity of the atmosphere. In absence of O$_3$, the average surface temperature of an Earth-like planet would be 7K cooler \citep{G_mez_Leal_2019}. Furthermore, O$_3$ serves as a filter for incoming UV radiation, safeguarding life on Earth's surface. Therefore, O$_3$ also plays a role in determining the potential habitability of an exoplanet.

To determine whether an exoplanet harbors life, scientists are keen on detecting biosignatures in its atmosphere. This detection can be achieved through direct imaging or transit spectroscopy. Scientists have considered using O$_3$ as a proxy to detect molecular oxygen (O$_2$) in the atmosphere of an exoplanet. However, a study conducted by \citet{Kozakis_2022} {has highlighted that the relationship between O$_3$ and O$_2$ is non-linear and depends on the type of stellar host. This makes it challenging to} accurately determine the amount of O$_2$ through O$_3$ measurements, especially when the UV flux of the host star is uncertain \citep{cooke2023degenerate}. Nonetheless, if we have knowledge about the host star's UV spectrum and certain other information, O$_3$ measurements might offer insights into the potential habitability of an exoplanet.

On Earth O$_3$ is primarily produced in the tropical stratosphere via the Chapman mechanism \citep{chapman1930theory}. During daytime, UV radiation breaks the O$_2$ molecule to form two O atoms which then reacts with O$_2$ molecules and a third body M through a three-body process to form O$_3$. The Brewer-Dobson circulation distributes O$_3$ in the stratosphere \citep{dobson1956origin, brewer1949evidence, newell1963transfer}. In this circulation pattern, air near the tropics rises and then moves toward the poles. The Brewer-Dobson circulation slightly differs between the two hemispheres due to variances in land and ocean structure in the Southern and the Northern Hemispheres \citep{article}.

This O$_3$ then either gets photodissociated by UV radiation to form O and O$_2$, or it reacts with O to form two O$_2$ molecules. O$_3$ also gets catalytically destroyed by NO$_x$, HO$_x$, Br, Cl, etc \citep{portmann2012stratospheric}. If X is a catalytic species, then the O$_3$ is catalytically destroyed by the following reactions: 
\begin{align*}
\qquad \text{X + O}_3 \rightarrow \text{XO + O}_2 \\
\qquad \text{XO + O} \rightarrow \text{X + O}_2 \\
\text{Net:} \quad \text{O}_3 + \text{O} \rightarrow 2\text{O}_2\\
\end{align*}

In 2016, three Earth-sized planets (TRAPPIST-1b, c, and d) were detected orbiting an ultra-cool M dwarf star known as TRAPPIST-1 or 2MASS J23062928 - 0502285, through observations made by the TRAPPIST (TRAnsiting Planets and PlanetIsimals Small Telescope) \citep{2016Natur.533..221G}. Subsequently, in 2017, Spitzer revealed the existence of four more exoplanets around TRAPPIST-1 (TRAPPIST-1e, f, g, and h), establishing it as the first system with seven Earth-sized planets \citep{2017Natur.542..456G}. All the planets in the TRAPPIST-1 system are closer to their host star than Mercury is to the Sun.

Future observations with JWST \citep{Gardner_2006} and new telescopes, such as ELT (Extremely Large Telescope) \citep{hook2009science} are expected to reveal more about small rocky exoplanets and their atmospheres. Looking ahead, future concepts like the Habitable Worlds Observatory \citep{national2021decadal}, which was proposed by combining two earlier concepts - the HabEx (Habitable Exoplanet Observatory Mission) \citep{gaudi2020habitable} and the LUVOIR (Large UV/Optical/IR Surveyor) \citep{theluvoirteam2019luvoir} -  are anticipated to directly image and characterize the atmospheres of Earth-like exoplanets. Hence, it becomes crucial to model rocky exoplanet atmospheres to aid in interpreting observations by extracting the atmospheric properties and explaining the underlying physical processes occurring in these exoplanetary atmospheres.

In previous studies, general circulation models (GCMs) have been employed to investigate the atmospheres of Earth-like planets, including studies designed to understand the atmospheric circulation and O$_3$ chemistry of tidally locked Earth-like exoplanets. For example, the TRAPPIST-1 Habitable Atmosphere Intercomparison (THAI) project \citep{turbet2022trappist,sergeev2022trappist} compared the results from four GCMs which included slab oceans for both dry and moist N$_2$-dominated and CO$_2$-dominated atmospheres. \citet{2018MNRAS.473.4672C} used the MITgcm to study the stratospheric circulation of a tidally locked ExoEarth scenario for TRAPPIST-1b, TRAPPIST-1d, Proxima Centauri b and GJ 667 C f. \cite{yates2020ozone} employed the Met Office Unified Model to explore the O$_3$ chemistry of the tidally locked exoplanet Proxima Centauri b around an M dwarf, while \cite{proedrou2016characterising} used the CESM1(WACCM) model to simulate the 3-D O$_3$ distribution of a tidally locked Earth-like planet around a Sun-like star. A recent study conducted by \citet{2023EGUGA..2515501B} utilized a slab ocean model of a tidally locked exoplanet around an M dwarf star with Proxima Centauri b parameters to study O$_3$ spatial distribution. They found that O$_3$ accumulates on the night side, demonstrating a day side-night side hemispheric asymmetry in O$_3$ distribution.

{Previous studies utilized slab ocean models to study atmospheric circulation on tidally locked exoplanets, resulting in a symmetric atmospheric circulation, as seen in \citet{2023EGUGA..2515501B}. However, many rocky exoplanets may not be completely covered by oceans, and it is unlikely that they will lack ocean dynamics. Furthermore, orography and landmass distribution play a significant role in shaping atmospheric circulation on Earth.} In our study we incorporated an Earth-like land-ocean structure to examine for the first time how orography might alter the atmospheric dynamics and chemistry of a {tidally locked Earth-like exoplanet modelled on TRAPPIST-1e.} Our aim is to characterize the possible O$_3$ distribution and the influence of atmospheric circulation on it. We focus on O$_3$ because it is affected by photochemistry, catalytic cycles, atmospheric transport, and has strong spectral features from the UV to the mid-infrared. Additionally, the presence of O$_2$, a potential biosignature, can be inferred from a detection of O$_3$.\\

\section{Methods} \label{sec:methods}

\subsection{Model Description and Setup}

For this study, we modeled TRAPPIST-1e using the Whole Atmosphere Community Climate Model version 6 (WACCM6)\citep{gettelman2019whole}. WACCM6 is an atmospheric model that operates as a configuration of the Community Earth System Model version 2 (CESM2). CESM2 is an Earth system model consisting of submodels simulating the atmosphere, ocean, land, sea ice, land ice, river runoff, and surface waves \citep{danabasoglu2020community}. The WACCM6 configuration comprises 70 vertical levels, starting from the surface at 1000 hPa and extending up to 140 km at 4.5 × 10$^{-6}$ hPa (lower thermosphere), with a horizontal resolution of 1.875{\textdegree} latitude × 2.5{\textdegree} longitude. 

The chemistry applied in WACCM6 is based on the Model of Ozone and Related Chemical Tracers (MOZART). MOZART serves as a global chemical transport model encompassing physical and chemical processes that span the troposphere, stratosphere, mesosphere, and lower thermosphere \citep{Emmons2020TheCM,conley2012description}. WACCM6 employs the Rapid Radiative Transfer Model (RRTMG) radiation code for solving longwave (lower energy radiation emitted from the Earth's surface and atmosphere) and shortwave (higher energy wavelengths such as UV, visible light, and a part of the near-infrared spectrum, associated with the host star) radiative transfer equations. {The longwave limit and the shortwave limit used here are 3.08–1000  $\mu$m and 0.2–12.2  $\mu$m, respectively.}

WACCM6 has been previously utilized to study the climate and O$_3$ chemistry of prehistoric Earth and Earth-like exoplanets.
For example, \citet{cooke2022revised} demonstrated that the total O$_3$ column at O$_2$ concentrations between 0.1 - 50\% of the present atmospheric level may have been lower than predicted by previous 1-D and 3-D models. {Possible reasons for this discrepancy include three-dimensional effects and transport, variations in boundary conditions, temperature structure and feedback mechanisms, seasonal cycles, calculations of absorption and scattering within the Schumann-Runge bands, and the condensation of water vapor through the tropical tropopause layer \citep{ji05300j}.} \citet{2023MNRAS.518..206C} showed how for an Earth-like exoplanet, observations and spectral signatures of chemical species could be affected by the line of sight, albedo, clouds, and chemistry. \citet{2023MNRAS.524.1491L} studied how changing the eccentricity of an Earth-like exoplanet affects the abundance and loss of water present in the atmosphere.

\begin{figure}[htb!]
    \centering
  \includegraphics[width=0.75\linewidth]{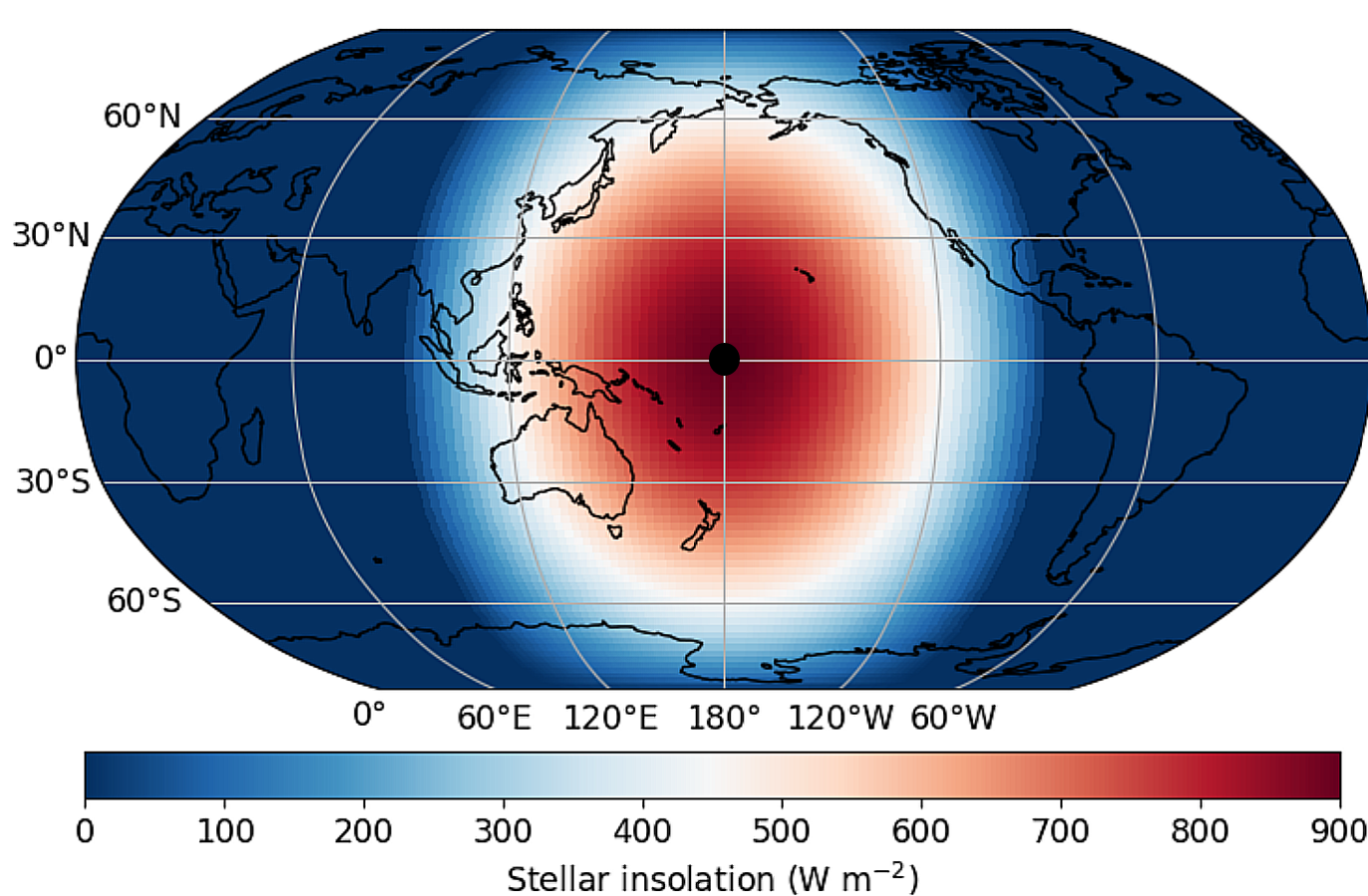}
  \caption{Stellar irradiation received by TRAPPIST-1e in our model. The black dot represent the substellar point which is fixed at 180{\textdegree} longitude and 0{\textdegree} latitude over the Pacific Ocean. The black outlines represent the land masses.}
  \label{fig:map}
\end{figure}

\begin{figure}[htb!]
    \centering
  \includegraphics[width=0.75\linewidth]{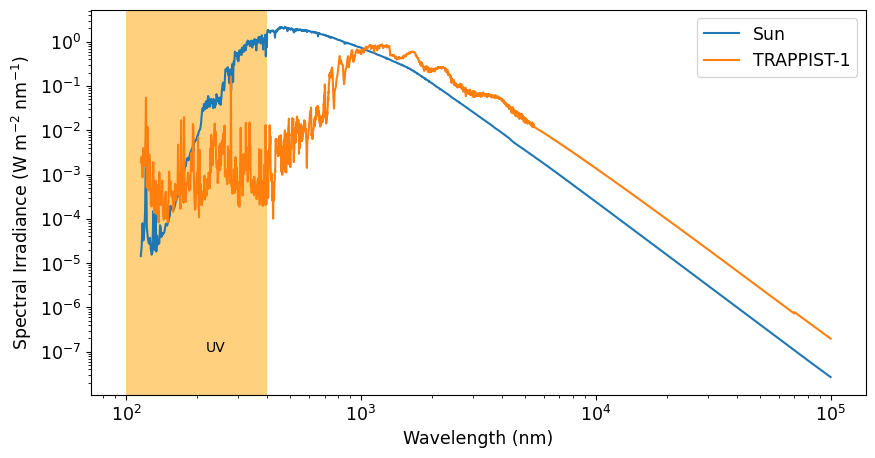}
  \caption{Stellar irradiance at top of the atmosphere of TRAPPIST-1e (orange) and Earth (blue). The orange shaded region represents the range of UV wavelengths.}
  \label{fig:spectra}
\end{figure}

The  model considered here used the BWma1850 compset\footnote{\url{https://docs.cesm.ucar.edu/models/cesm2/config/2.1.3/compsets.html}} of WACCM6, which included a pre-industrial Earth-like atmosphere and orography, and was modified to allow for synchronous rotation.\footnote{\url{https://github.com/exo-cesm/CESM2.1.3/tree/main/Tidally_locked_exoplanets/cases}} The model composition was with O$_2$ set to present atmospheric level (PAL), i.e., 21\% by volume, and N$_2$ with 78\% by volume. The volume mixing ratios of CH$_4$ (0.8 ppmv), CO$_2$ (280 ppmv), N$_2$O (270 ppbv), and H$_2$ (500 ppbv), are fixed at the surface. We ran the simulation for 300 years, of which we utilised the last 40 years of data in order to eliminate any effects associated with the model adjusting to tidally locked conditions. The substellar point was fixed at 180{\textdegree} longitude and 0{\textdegree} latitude over the Pacific Ocean (see Figure \ref{fig:map}). Table \ref{table} shows the parameters used to model TRAPPIST-1e. For this study, we have used a stellar spectrum based on the work of \citet{peacock2019predicting} who modelled the stellar energy distribution of TRAPPIST-1 and produced model 1A, 2A and 2B of which we used model 1A, which best matched the TRAPPIST-1 Ly$\alpha$ reconstruction from \citet{bourrier2017reconnaissance}. Figure \ref{fig:spectra} compares the top-of-the-atmosphere irradiation of TRAPPIST-1e and Earth.\\

\begin{table}
\centering
\begin{tabular}{ c  c  c }
\hline
Parameter & Units & Value \\
\hline\hline
Semimajor axis & au & 0.029 \\
\hline
Orbital period & Earth days & 6.1 \\
\hline
Rotation period & Earth days & 6.1 \\
\hline
Obliquity &  & 0 \\
\hline
Eccentricity &  & 0 \\
\hline
Instellation & W m$^{-2}$ & 900 \\
\hline
Planet radius & km & 5797 \\
\hline
Gravity & m s$^{-2}$ & 9.14 \\
\hline
\end{tabular}
\caption{Planetary parameters used for model of TRAPPIST-1e \citep{Grimm}.}
\label{table}
\end{table}

\subsection{Data Analysis}

For our analysis, we calculated the time average of the 40 years of data and examined various parameters, including surface temperature, horizontal wind velocities ($u$ and $v$), vertical wind velocity ($w$), reaction rates, and volume mixing ratios of O$_3$, O$_2$, OH, HO$_2$, NO, NO$_2$, Br, and Cl. To calculate the concentration of chemical species in units of molecules m$^{-3}$, we multiplied the air number density by the volume mixing ratio of the respective chemical species.

We calculated the O$_x$ production rate using the formula $J_{O_2}$*[$O_2$], where $J_{O_2}$ is the oxygen photolysis rate constant and [$O_2$] is the O$_2$ concentration. The catalytic destruction rates of O$_3$ by various chemical species (X) were calculated using the formula $k$*[X]*[O$_3$], where $k$ is the rate constant for X + O$_3$ reactions, [X] is the concentration of the catalytic species, and [O$_3$] is the concentration of O$_3$. In this context, X can be OH, HO$_2$, NO, NO$_2$, Br, or Cl. 

To study the horizontal wind circulation we used the Helmholtz wind decomposition, which has been previously used to study the atmospheric circulations of Earth, tidally locked terrestrial exoplanets and hot Jupiters \citep{Hammond_2021}. The Helmholtz wind decomposition breaks down the total horizontal wind into rotational ($\boldsymbol{u_r}$) and divergent components ($\boldsymbol{u_d}$), allowing us to study the winds responsible for the overall circulation in more detail. For a tidally locked planet, the rotational flow consists of two parts: an equatorial jet (zonal-mean rotational component) which is a narrow belt of winds moving around the equator of a planet and stationary waves (eddy rotational component). The stationary waves drive the equitorial jet and can accelerate it resulting in a superrotating jet which moves faster than the planet's rotation rate \citep{Showman_2013}.  The divergent flow consists of a global overturning circulation which is a result of thermally driven circulation where the air rises on the day side of the planet and sinks on the night side. Mathematically, the Helmholtz wind decomposition divides the total circulation given by $\boldsymbol{u}$ = ($u$,$v$) into the two aforementioned components \citep{Hammond_2021}:

\begin{equation}
    \label{eq:equation1}
    \boldsymbol{u} = \boldsymbol{u_d} + \boldsymbol{u_r} =  -\boldsymbol{\nabla}\chi + \boldsymbol{\mathrm{k}} \times \boldsymbol{\nabla}\psi
\end{equation}

\begin{equation} 
    \label{eq:equation2}
    \nabla{^2}\chi = \delta
\end{equation}
\begin{equation} 
    \label{eq:equation3}
    \nabla{^2}\psi = \zeta
\end{equation}

Here, $\boldsymbol{u}$ represents the horizontal wind vector, where $u$ and $v$ denote the zonal (along latitude) and meridional (along longitude) velocities, respectively. The velocity potential function, $\chi$, is derived from the divergence ($\delta$ - Equation \ref{eq:equation2}), while the streamfunction, $\psi$, is derived from the vorticity ($\zeta$ - Equation \ref{eq:equation3}) of the wind.

To examine the meridional overturning circulation, we used the meridional mass streamfunction. This is commonly used to study atmospheric or oceanic circulation patterns and helps in visualizing mass transport in the meridional plane (North-South direction). Mathematically, the relationship between mass streamfunction $\psi$ and both vertical velocity ($w$) and meridional velocity ($v$) is expressed by the equations \citep{2020GMD....13.3337S}

\begin{equation} 
    \label{eq:equation4}
    \frac{1}{R} \frac{\partial \psi}{\partial \theta} = w,  \quad \frac{\partial \psi}{\partial z} = -v
\end{equation}

 Here, $\theta$ is the latitude in radians, $z$ is the altitude and $R$ is the planet's radius.\\

\section{Results and Discussion} \label{sec:Results and Discussion}

\subsection{Ozone Distribution}

\begin{figure}[btp]
    \centering
  \includegraphics[width=\linewidth]{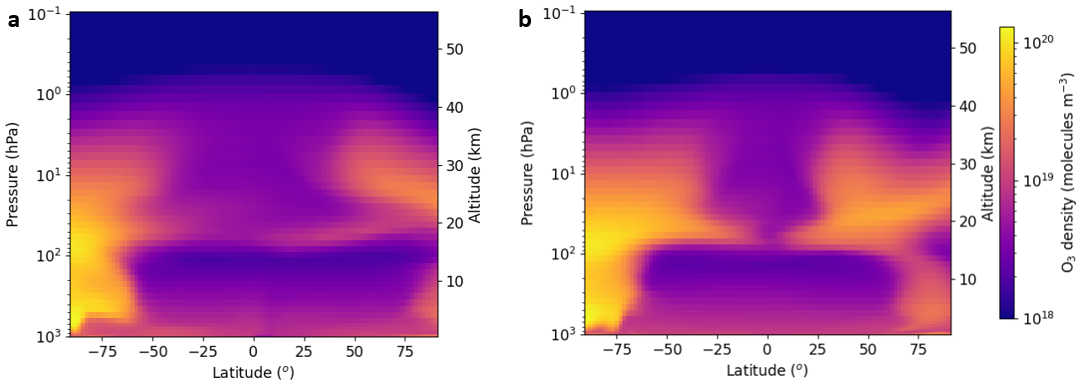}
\caption{Cross section of the O$_3$ number density on the meridional plane passing through the a) substellar point and b) antistellar point for our TRAPPIST-1e model.}
\label{fig:ozone}
\end{figure}

Figure \ref{fig:ozone} a) and b) show the {O$_3$ concentration on the meridional plane passing through the substellar point and the antistellar point for our TRAPPIST-1e model. We consider the profiles at the substellar point and the antistellar point to be representative of the ‘day side’ and the ‘night side’, respectively.} O$_3$ is predominantly present in the lower atmosphere, at pressures $>$ 10 hPa (below $\sim$ 30 km) and O$_3$ concentrations are higher near the poles than near the equator. The highest O$_3$ concentrations are found near the South Pole between pressures 600 and 70 hPa (between $\sim$5 and $\sim$15 km). Further, there is little difference in O$_3$ concentration between the ‘day side’ and the ‘night side’, much less than the North-South asymmetry, suggesting that the O$_3$ produced on the day side {gets mixed across longitudes by the horizontal winds.}

\begin{figure}[btp]
    \centering
    \centering
  \includegraphics[width=\linewidth]{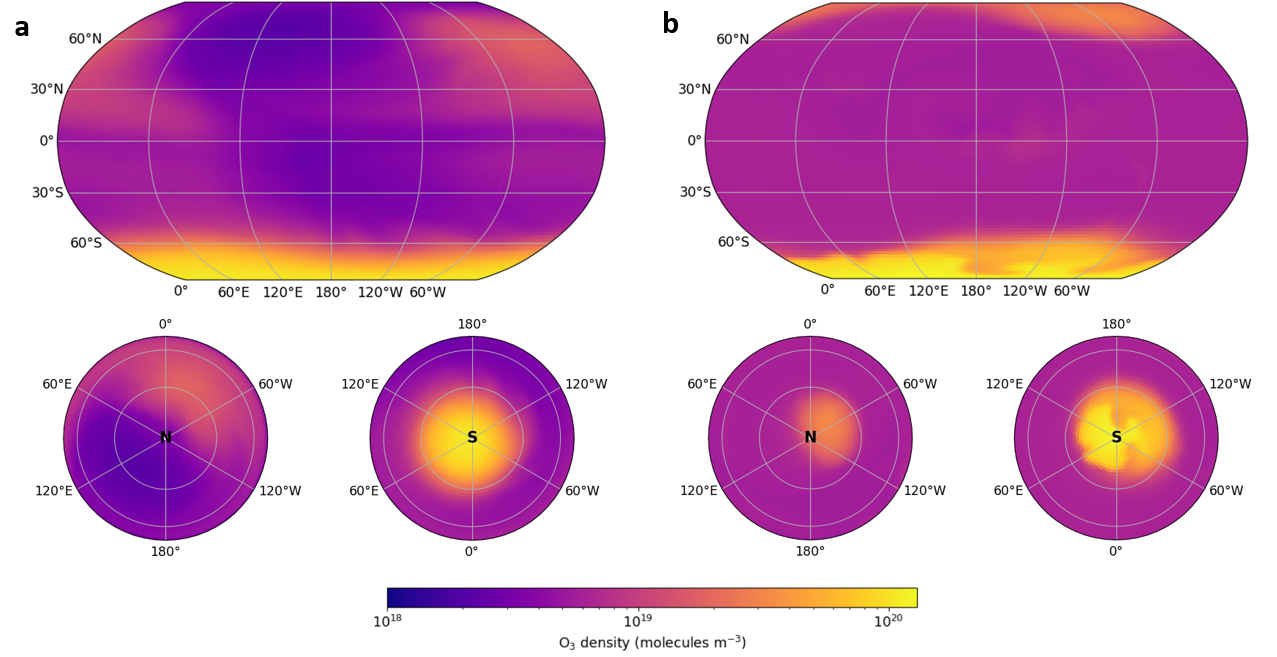}
\caption{Horizontal slices of the O$_3$ number density for our TRAPPIST-1e model at a pressure (altitude) of a) 103 hPa ($\sim$15 km) and b) 609 hPa ($\sim$5 km). The substellar point is at 180{\textdegree} longitude and 0{\textdegree} latitude. N represents the North Pole and S represents the South Pole. Lower panels show the polar projections.}
\label{fig:ozone 2}
\end{figure}

Figure \ref{fig:ozone 2} a) and b) display the O$_3$ concentrations at pressures 103 hPa ($\sim$15 km) and 609 hPa ($\sim$5 km) respectively. At both of these pressures, the North-South asymmetry in O$_3$ distribution is clearly visible. {It also demonstrates that the difference in O$_3$ concentration between the day side and the night side is significantly smaller compared to the North-South difference (see also Figure 9 (SSPO) in \citet{sainsburymartinez2024landmassdistributioninfluencesatmospheric}).} {On Earth, O$_3$ produced in the stratosphere near the equator is transported poleward and downward by the Brewer-Dobson circulation, which moves O$_3$-rich air towards the poles. This circulation plays a key role in maintaining a relatively symmetric O$_3$ distribution between the Northern and Southern Hemispheres, though some asymmetries arise due to geographic features, seasonal variations, and polar atmospheric dynamics.} The majority of O$_3$ is located between pressures 300 and 10 hPa. Between pressures 30 and 10 hPa, O$_3$ is densely packed around low latitudes, while between pressures 300 and 30 hPa, O$_3$ is more abundant around the poles (see \citet{Ejzak_2007}). Therefore, the vertical structure and the latitudinal distribution of O$_3$ in our TRAPPIST-1e model differs significantly from that of Earth.

\begin{figure}[btp]
    \centering
  \includegraphics[width=0.55\linewidth]{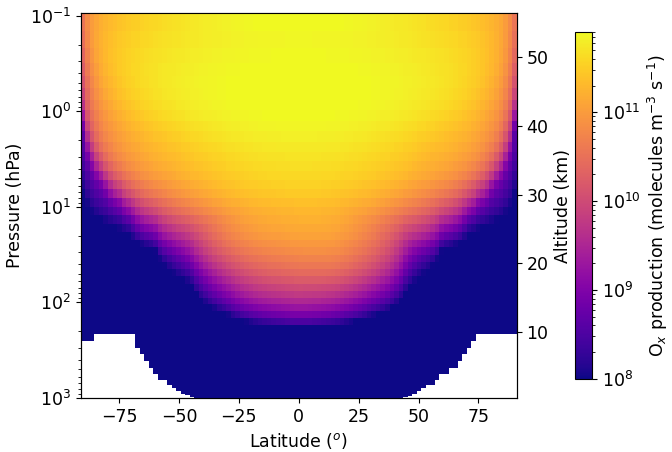}
  \caption{Cross section of O$_x$ production rate on the meridional plane passing through the substellar point for our TRAPPIST-1e model. The white region on the plot indicates zero production rate.}
  \label{fig:Ox production}
\end{figure}

Figure \ref{fig:Ox production} shows the O$_x$ production rate (O + O$_3$ production rate) on the meridional plane passing through the substellar point. The O$_x$ production rate is symmetric, with peak production occurring in the upper atmosphere (at pressures $<$ 10 hPa or at altitudes above $\sim$30 km) due to the large amount of UV radiation received here. UV radiation does not penetrate the lower atmosphere because it becomes denser, absorbing most of the incoming UV radiation, and it is weakest at high latitudes due to geometric effects. Hence, in the lower atmosphere, there is less O$_x$ production near the equator and no production at higher latitudes (as seen in figure \ref{fig:Ox production}, where white areas indicate regions of zero production).
When comparing the pattern of O$_3$ concentration and O$_x$ production rate, it is evident that the region of peak O$_x$ production rate does not coincide with the region of peak O$_3$ concentration. There is relatively low O$_3$ concentrations where the O$_x$ production is high, whereas the region with the highest O$_3$ concentration (near the South Pole between pressures 600 and 70 hPa) has either low or no O$_x$ production. This indicates that O$_3$ is primarily produced high up in the atmosphere and either all the O$_3$ is transported from the upper atmosphere to the lower atmosphere, or some of it is transported while the remainder is catalytically destroyed. However, this does not explain the North-South asymmetry in O$_3$ concentration. To explain this asymmetry, we need to examine if there are asymmetries in catalytic species distribution and atmospheric circulation.

\begin{figure}[btp]
    \centering
  \includegraphics[width=0.45\linewidth]{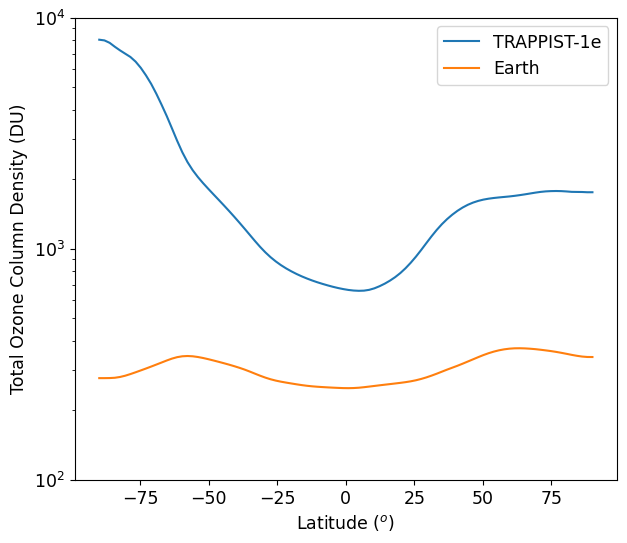}
  \caption{Zonal mean latitudinal variation of total ozone column density for our TRAPPIST-1e model (blue) and Earth (orange).}
  \label{fig:TOC}
\end{figure}

Total ozone column (TOC) density is reported in Dobson Units (DU), where 1 DU is equivalent to a 10 $\mu$m thick layer of pure O$_3$ at 273 K and 1 atm pressure. For Earth the values of TOC are 200-250 DU near the South Pole and 300-350 DU near the North Pole.\footnote{\url{https://ozonewatch.gsfc.nasa.gov/SH.html} - Date: 24/08/2023}
Figure \ref{fig:TOC} shows the latitudinal variation in zonal mean TOC density for both our TRAPPIST-1e model and Earth. From this figure, we clearly observe the North-South asymmetry in O$_3$ distribution in our TRAPPIST-1e model, whereas O$_3$ on Earth is roughly symmetrically distributed about the equator. {Note that, we find that the } total O$_3$ concentration in our TRAPPIST-1e model is much higher compared to that of Earth. The TOC density in our TRAPPIST-1e model near the South Pole is 8000 DU, which is around 28 times that of Earth ($\sim$250 DU), and near the North Pole, it is approximately 2000 DU, which is 7 times that of Earth ($\sim$300 DU). There is slight asymmetry in TOC on Earth with slightly higher TOC near the North Pole as compared to the South Pole, but for our TRAPPIST-1e model this asymmetry is reversed, and significantly larger. 

\subsection{Role of Catalytic Species}

\begin{figure}[btp]
\gridline{\fig{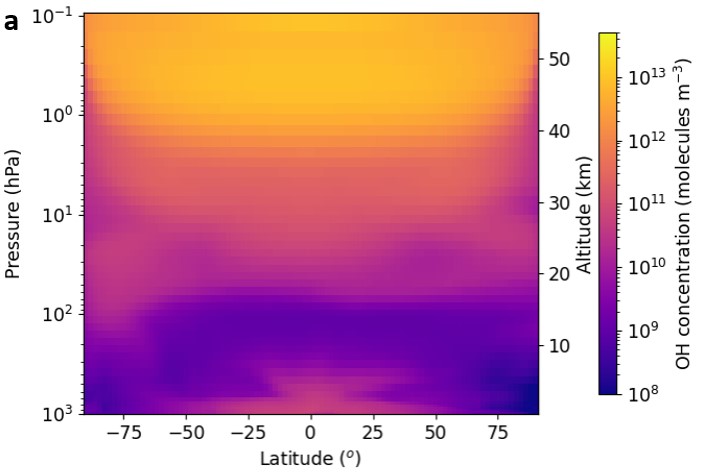}{0.45\textwidth}{}
          \fig{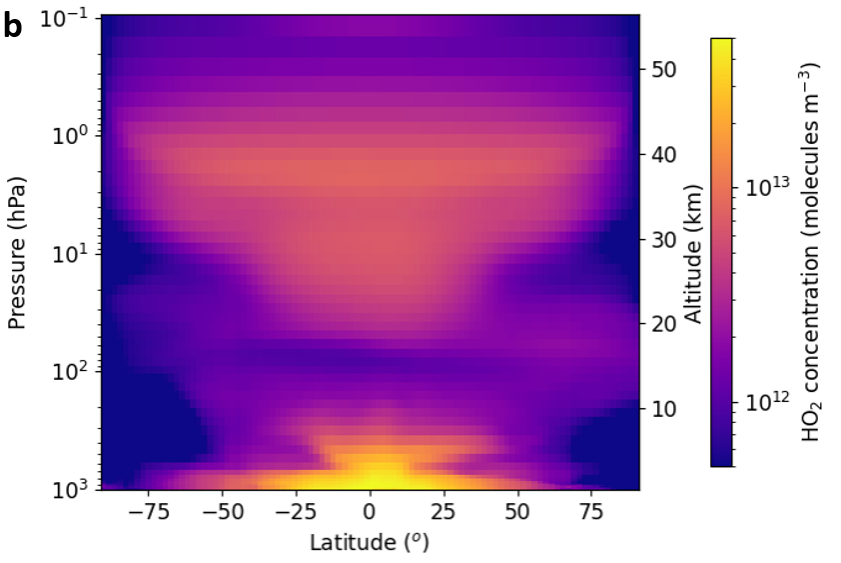}{0.45\textwidth}{}}
\gridline{\fig{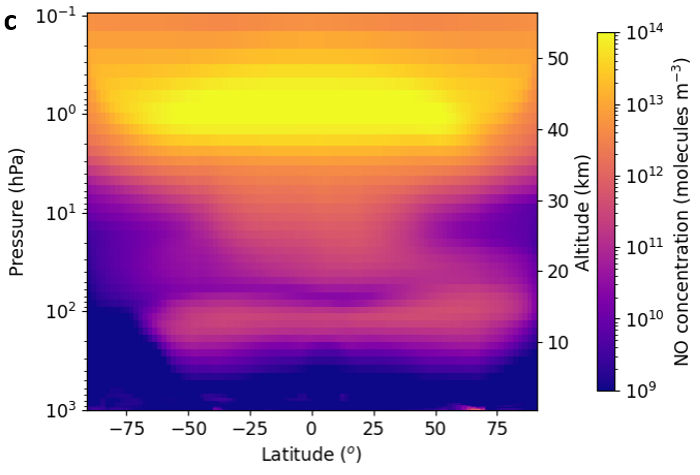}{0.45\textwidth}{}
          \fig{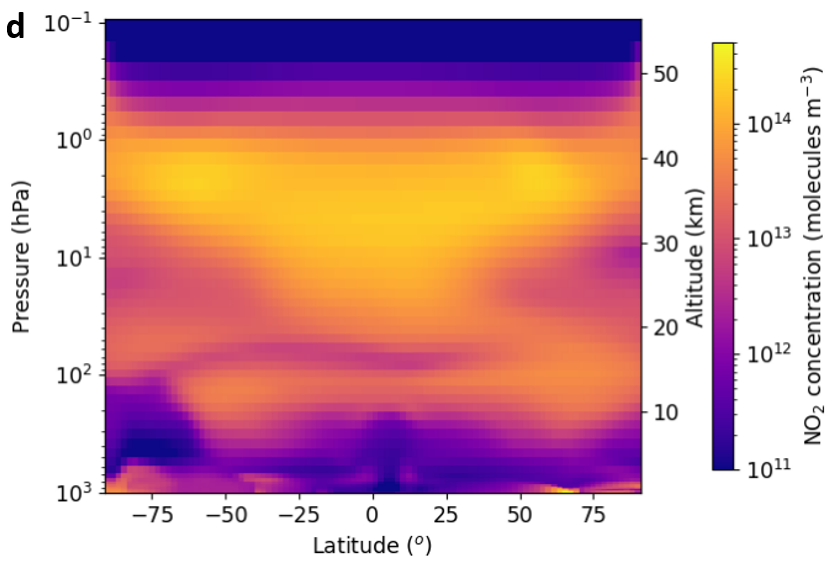}{0.45\textwidth}{}}
\gridline{\fig{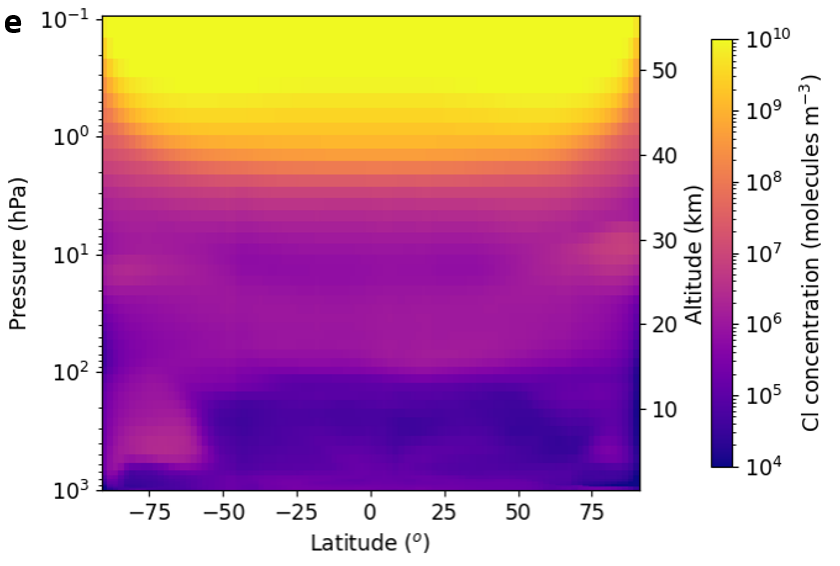}{0.45\textwidth}{}
          \fig{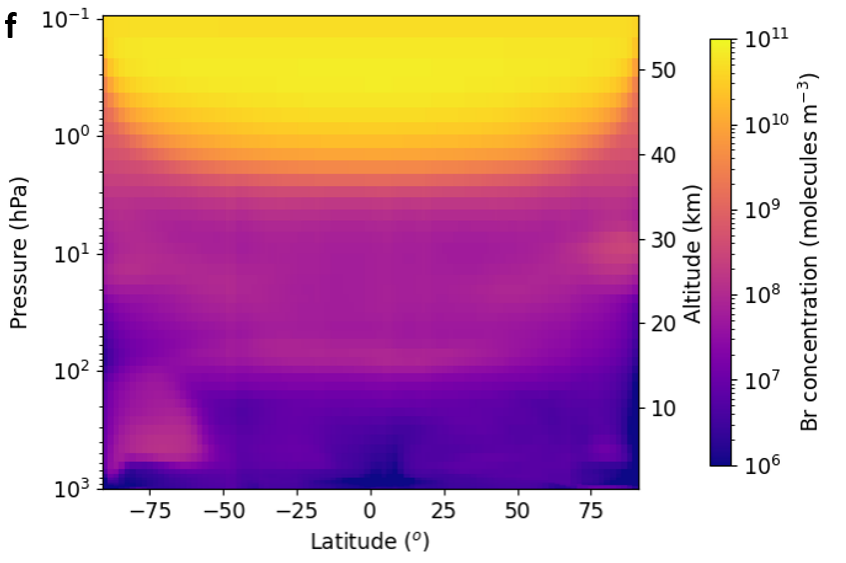}{0.45\textwidth}{}}
\caption{Catalytic species concentration on the meridional plane passing through the substellar point for a) OH, b) HO$_2$, c) NO, d) NO$_2$, e) Cl and f) Br for our TRAPPIST-1e model. The catalytic species are roughly symmetrically distributed and their concentrations are lower than the O$_3$ concentrations.}
\label{fig:Catalyst}
\end{figure}

Figure \ref{fig:Catalyst} shows the distribution of catalytic species (OH, HO$_2$, NO, NO$_2$, Cl, and Br) on the meridional plane passing through the substellar point for our TRAPPIST-1e model. For regions at pressures $>$ 10 hPa (below $\sim$30 km), our initial hypothesis, after observing the symmetric O$_x$ production and the asymmetric distribution of O$_3$, was that there might be an uneven distribution of catalytic species. Such an imbalance could potentially be responsible for O$_3$ depletion being more pronounced near the North Pole compared to the South Pole. However, our analysis suggests that this is not the case, with the distribution of catalytic species being roughly symmetric and the concentrations low when compared to the O$_3$ concentration. In the case of Cl (Figure \ref{fig:Catalyst} e) and Br (Figure \ref{fig:Catalyst} f) we find that at pressures $>$ 100 hPa the concentrations of these species is slightly higher near the South Pole as compared to the North Pole, since this is aligned with the peak in O$_3$ concentration, it cannot explain the high concentrations of O$_3$ we find at pressures $>$ 10 hPa near the South Pole.   

\begin{figure}[btp]
    \centering
  \includegraphics[width=0.55\linewidth]{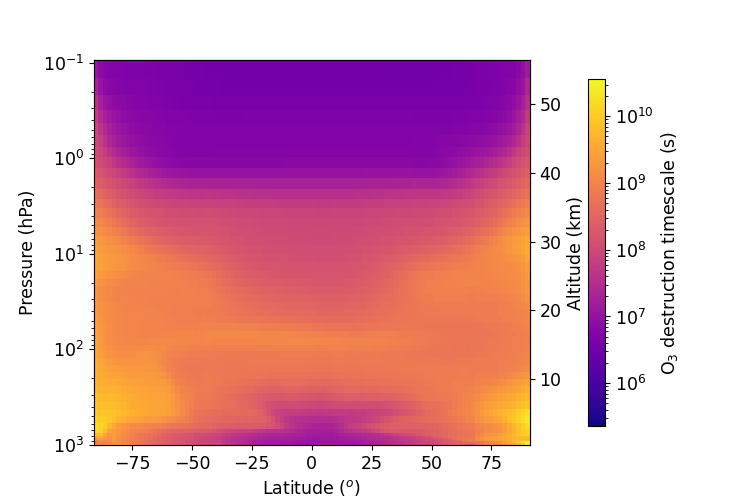}
  \caption{Time taken by the catalytic species to destroy O$_3$ on the meridional plane passing through the substellar point for our TRAPPIST-1e model.}
  \label{fig:O3 destruction time}
\end{figure}

Figure \ref{fig:O3 destruction time} shows the time taken by the catalytic species to destroy O$_3$ on the meridional plane passing through the substellar point. We observe that in regions with the highest concentrations of O$_3$, catalytic species take longer to destroy O$_3$ compared to regions where O$_3$ is produced. {The higher concentration of O$_3$ means there is more O$_3$ to destroy, and the concentration of catalytic species is low relative to O$_3$, leading to a slower overall destruction rate. Furthermore, the replenishment of O$_3$ through production on the day side and its transport to the night side further slows the catalytic destruction process.} This observation suggests that at pressures $>$ 10 hPa, catalytic destruction occurs at a slower rate, allowing atmospheric circulation to influence the distribution of O$_3$. As such, this suggests that catalytic species may not be the primary factor driving the asymmetry in the distribution of O$_3$. Therefore, we next look at the transport of O$_3$ throughout the atmosphere to see if this might have a more significant effect.

\subsection{Role of Atmospheric Circulation}

\begin{figure}[tbp]
\gridline{\fig{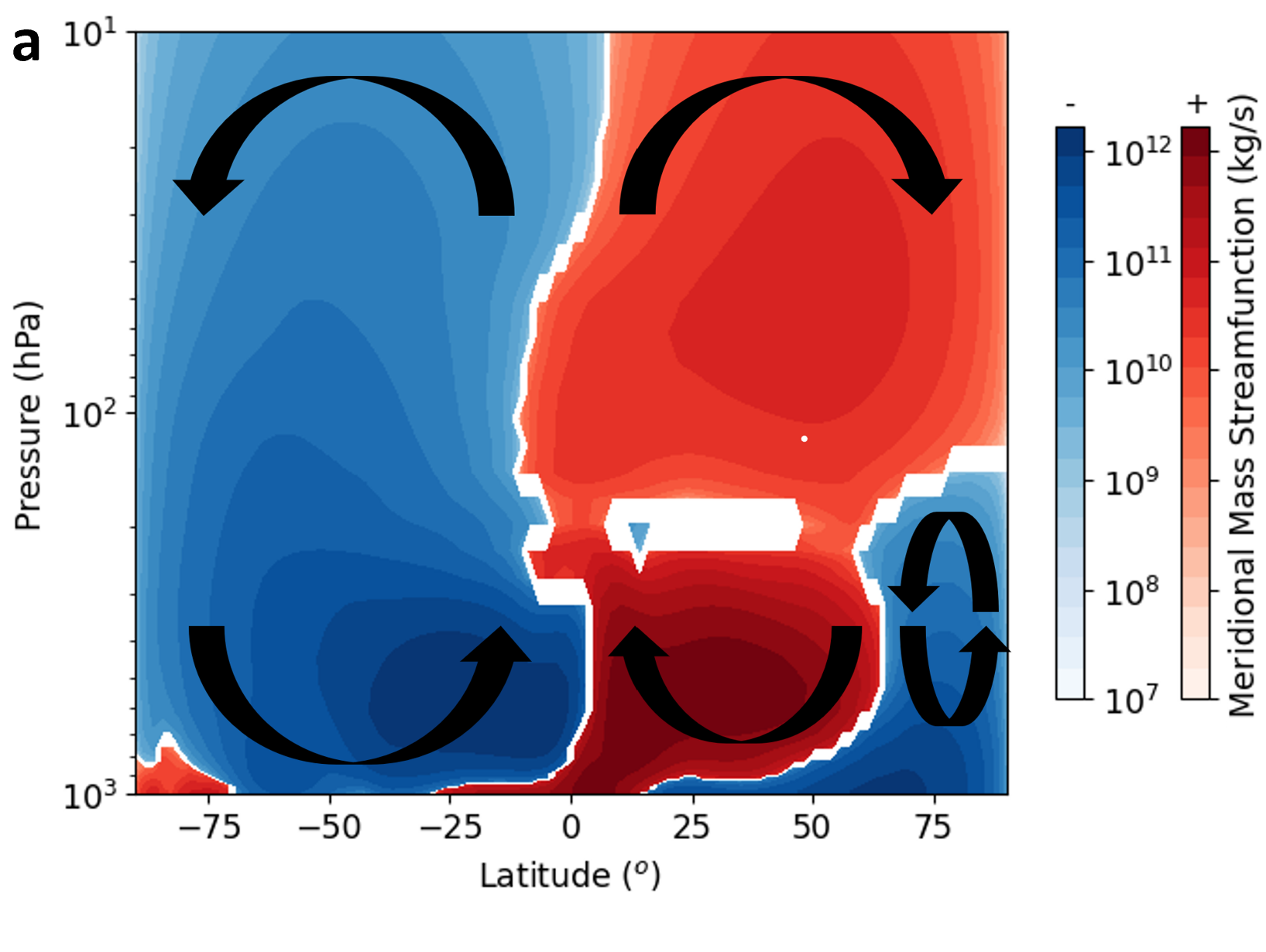}{0.45\textwidth}{}
          \fig{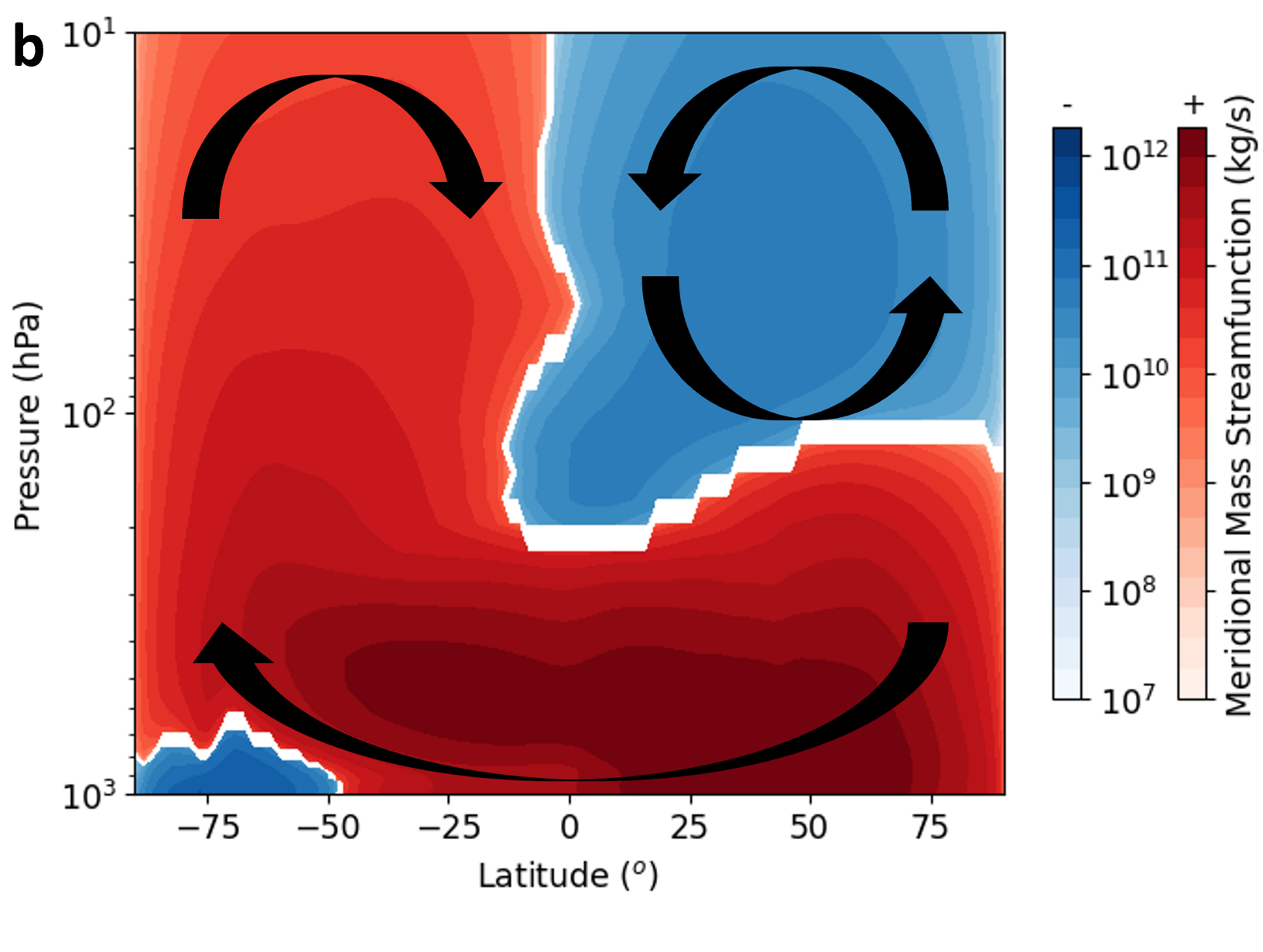}{0.45\textwidth}{}}
\caption{Meridional mass streamfunction on the meridional plane passing through the a) substellar point and the b) antistellar point for our TRAPPIST-1e model. The positive values (red) indicates clockwise circulation and the negative values (blue) indicates anticlockwise circulation. The values are in log-scale.}
\label{fig:MSF}
\end{figure}

Figure \ref{fig:MSF} a) and b) shows the meridional mass streamfunction on a plane passing through the substellar point and the antistellar point, respectively. {We consider the profile at the substellar point to be representative of the ‘day side’ and the profile at the antistellar point to be representative of the ‘night side’.} Here, red indicates clockwise transfer of mass and blue indicates anticlockwise transfer of mass. On the ‘day side’, without orography we would expect a symmetric circulation with upwelling at the equator, meridional flow towards the poles, downwelling near the poles and meridional flow from the poles to the equator (see \cite{sergeev2022trappist}). In our simulation, we find that there is an asymmetry in this circulation: near the North Pole the {clockwise} downwelling is interrupted at 200 hPa and does not reach the surface, rather there is a {clockwise} meridional transport from the North Pole towards mid latitudes, {clockwise} downwelling at mid latitude and then {clockwise} meridional flow from mid latitude towards the equator (Figure \ref{fig:MSF} a). {The formation of a small anticlockwise (blue) circulation cell near the North Pole breaks the symmetry.} Similarly, on the ‘night side’, without orography we would expect to find upwelling near the poles, meridional transport from the poles to the equator, downwelling at the equator and meridional transport from the equator towards the poles near surface, thus closing the global overturning circulation (see \cite{sergeev2022trappist}). However, figure \ref{fig:MSF} b) reveals there is an asymmetry in this circulation. Near the North Pole there are two circulation cells: at pressures $<$ 200 hPa, there is {anticlockwise} upwelling from the North Pole, {anticlockwise} meridional transport from the North Pole to the equator, {anticlockwise} downwelling near the equator until pressure 200 hPa, and {anticlockwise} meridional transport back to the North Pole. At pressures $>$ 200 hPa, there is {clockwise} downwelling near the North Pole, {clockwise} meridional transport from the North Pole, across the equator, to the South Pole, and then {clockwise} upwelling near the South Pole.

{Overall}, both on the ‘day side’ and the ‘night side’, we find an asymmetry in the meridional overturning circulation in the Northern Hemisphere. This asymmetry is caused by the presence of Earth-like orography. Both on the day side and the night side, the Northern Hemisphere has a higher fraction of landmass compared to the Southern Hemisphere. This landmass reshapes the winds, {leading to the formation of a small anticlockwise cell on the ‘day side’ and a large clockwise cell on the ‘night side’,} resulting in an asymmetric meridional overturning circulation {in the Northern Hemisphere.}

\begin{figure}[btp]
\centering
\gridline{\fig{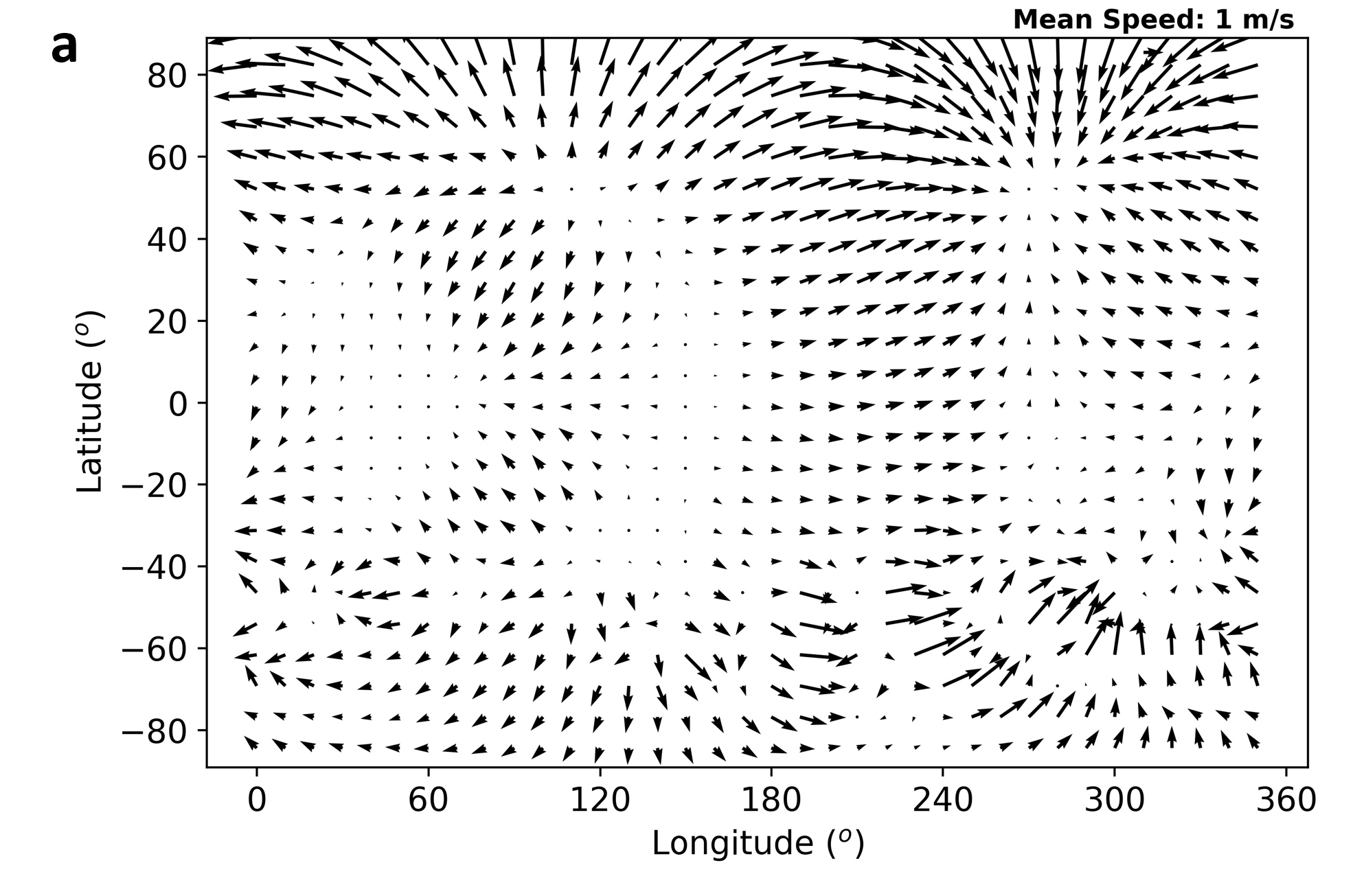}{0.45\textwidth}{}
          \fig{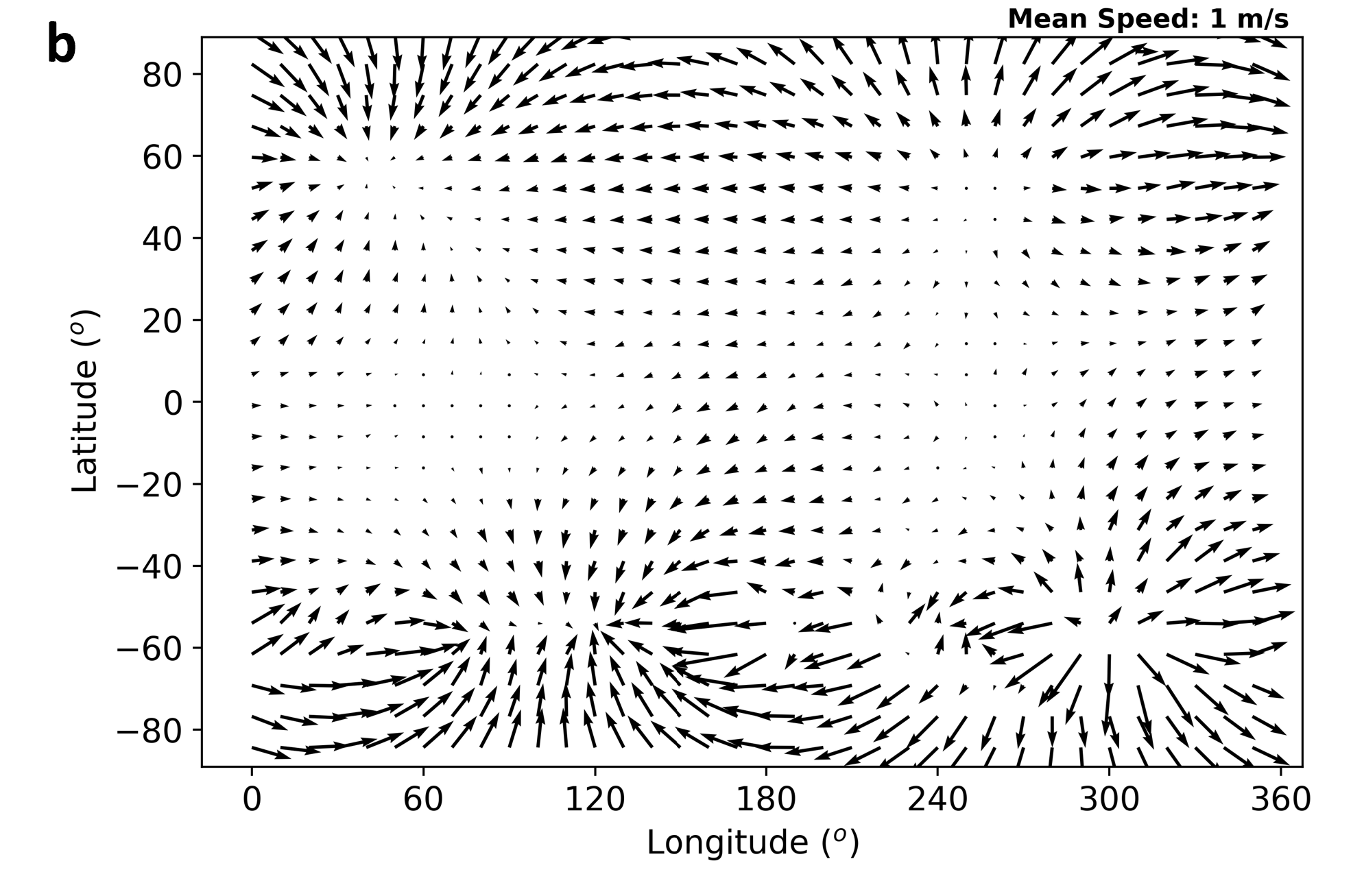}{0.45\textwidth}{}}
\gridline{\fig{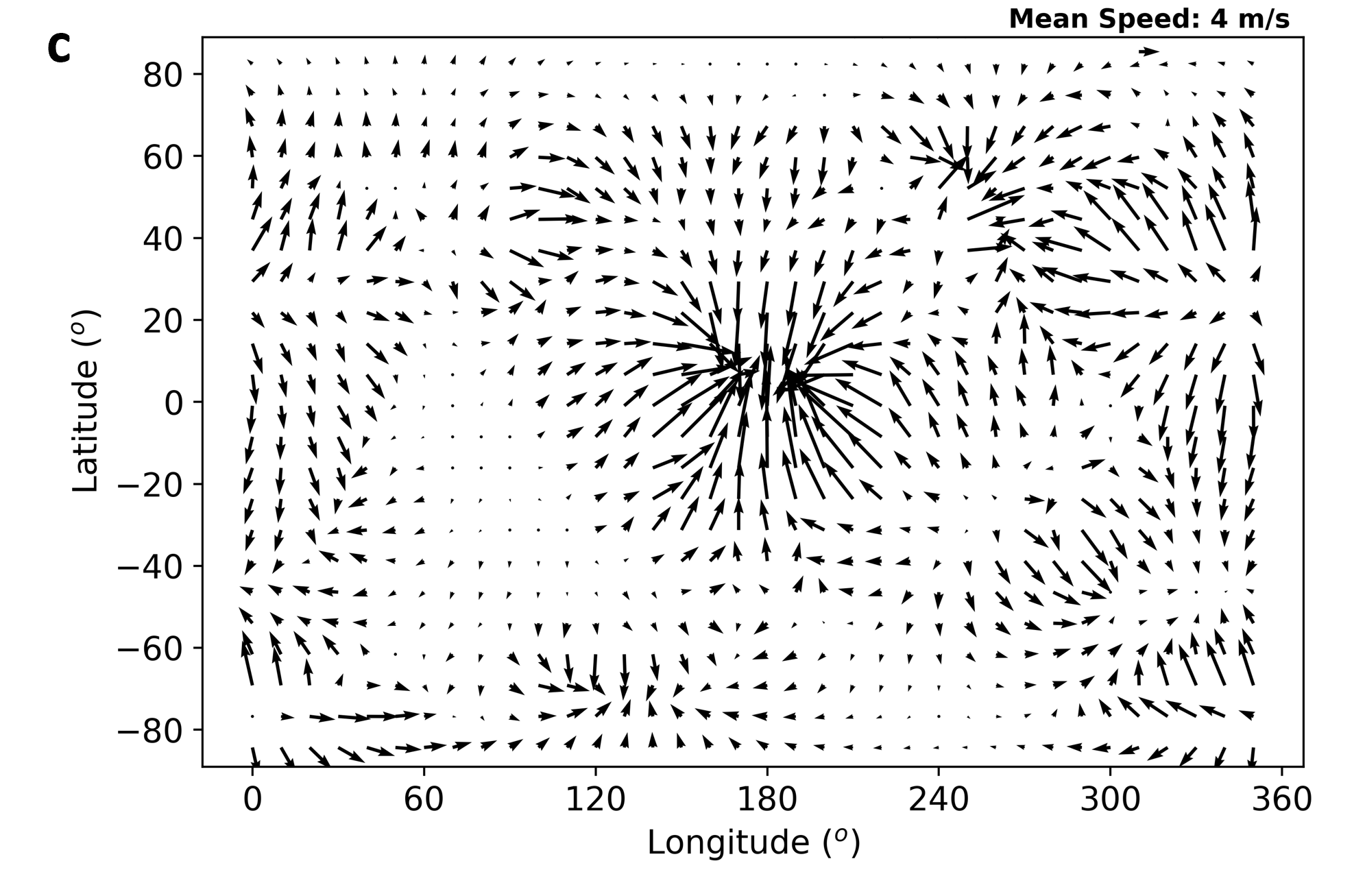}{0.45\textwidth}{}}    
\caption{Helmholtz divergent winds at pressures a) 20 hPa, b) 103 hPa and c) 800 hPa. Arrows represents winds velocity vectors and point in the direction of flow, with the length of the arrows representing the magnitude of the wind speed (with the mean wind speed shown in the top right corner of each plot).}
\label{fig:Helmholtz_d}
\end{figure}

\begin{figure}[btp]
\centering
\gridline{\fig{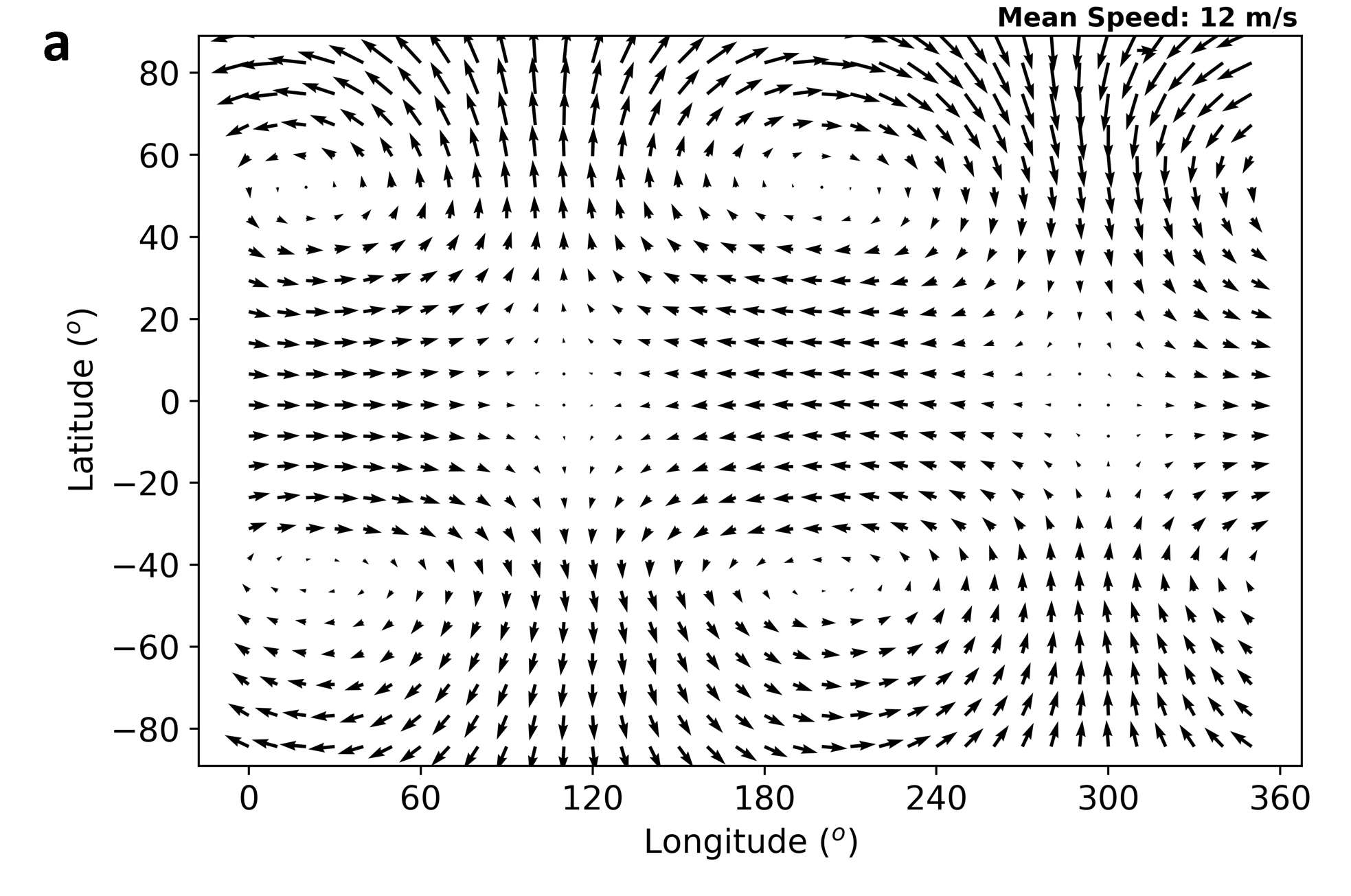}{0.45\textwidth}{}
          \fig{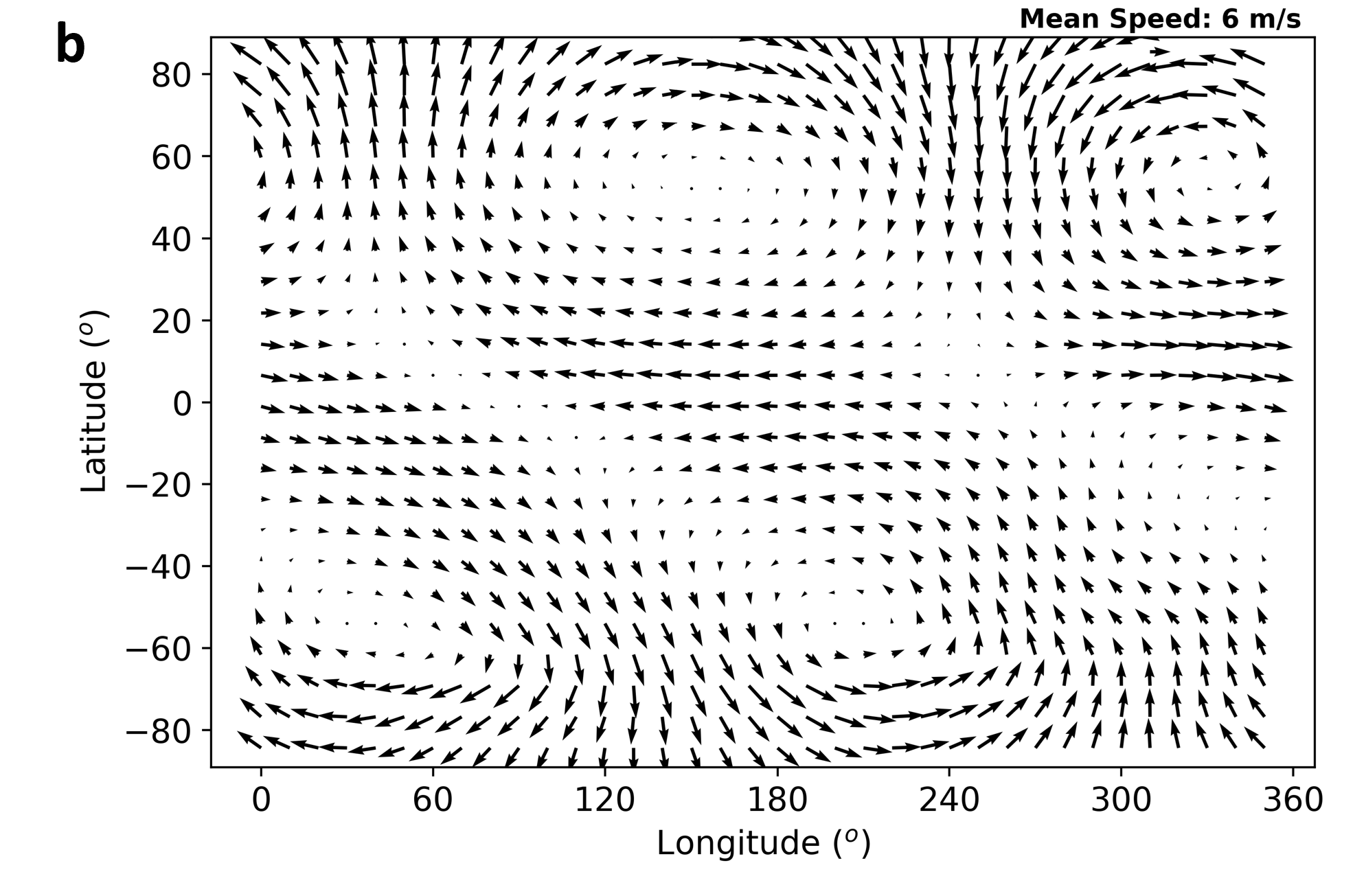}{0.45\textwidth}{}}
\gridline{\fig{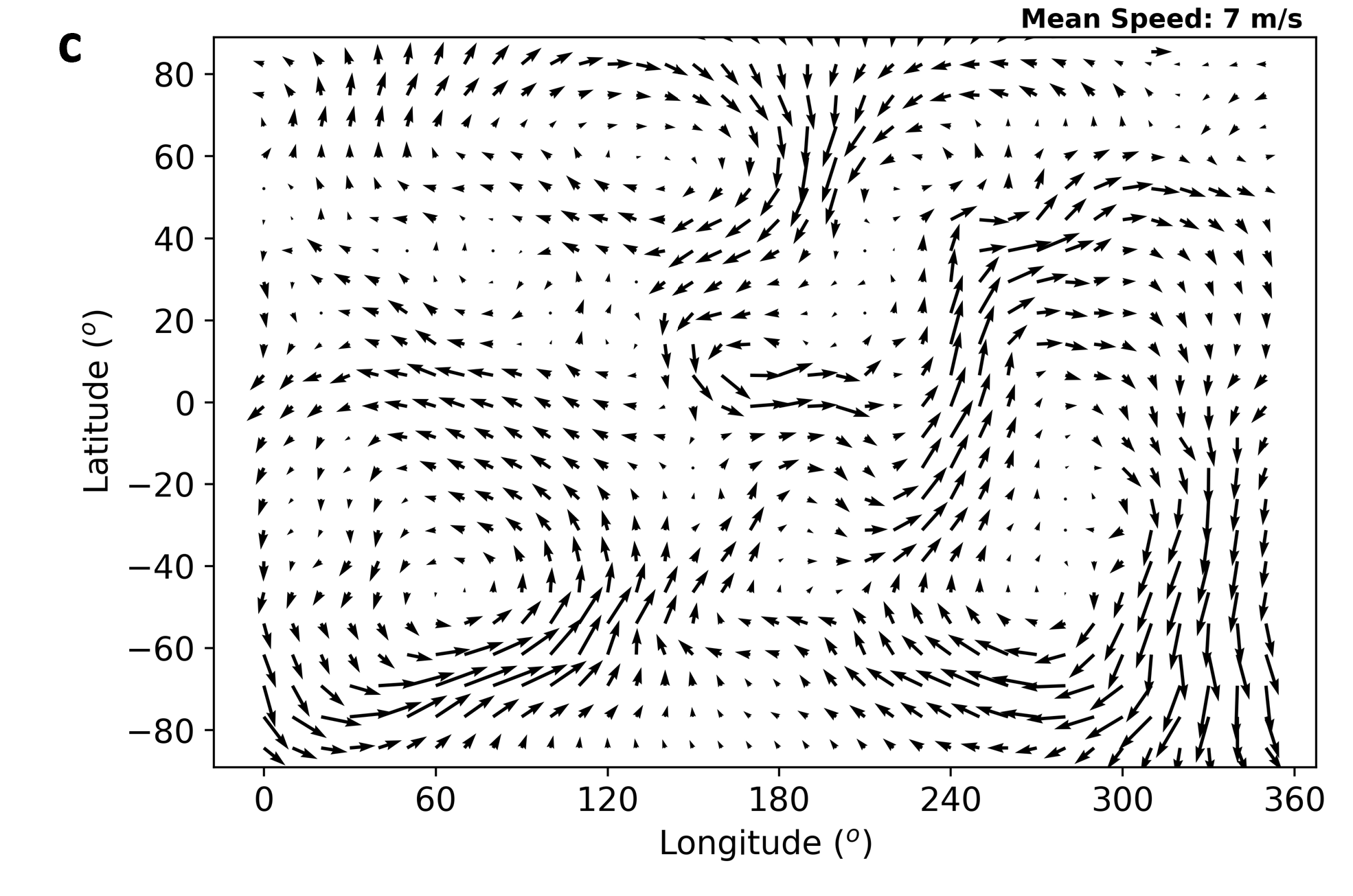}{0.45\textwidth}{}}    
\caption{Helmholtz rotational eddy winds at pressures a) 20 hPa, b) 103 hPa and c) 800 hPa. Arrows represent winds velocity vectors and point in the direction of flow, with the length of the arrows representing the magnitude of the wind speed (with the mean wind speed shown in the top right corner of each plot).}
\label{fig:Helmholtz_r}
\end{figure}

{To visualize the complex horizontal winds, we use Helmholtz wind decomposition, which breaks down the total horizontal wind into its rotational (divergence-free) and divergent (rotation-free) components.} Figure \ref{fig:Helmholtz_d} and Figure \ref{fig:Helmholtz_r} show the divergent component and the eddy rotational component of the horizontal winds in the stratosphere (20 hPa), near the tropopause (103 hPa), and near the surface (800 hPa). When comparing the mean wind speeds of both components at each pressure level, we find that the eddy rotational component has a higher mean wind speed than the divergent component. This implies that the eddy rotational component plays a major role in meridional transport.

We can explain the O$_3$ concentrations North-South asymmetry using both the meridional overturning circulation and the Helmholtz wind decomposition. O$_3$ formed on the day side between pressures 200 and 10 hPa (between $\sim$10 and $\sim$30 km) is evenly transported {towards} the poles by meridional flow and then from the day side to the night side by the rotational component of horizontal winds, which includes both the eddy and zonal-mean components. On the night side, in the lower atmosphere (at pressures $>$ 100 hPa), the meridional flow (see Figure \ref{fig:Helmholtz_r} c) forces large amounts of O$_3$ from the North Pole to the South Pole. {The North-South asymmetry in the meridional flow, near the surface, on both the ‘night side’ and the ‘day side’ occurs due to the presence of Earth-like orography. Land-ocean boundaries and orography significantly influence the near-surface winds, including slowing and redirecting them. For more details, see \citet{sainsburymartinez2024landmassdistributioninfluencesatmospheric}. This orographic effect drives a near-surface North-to-South flow on the night side, disrupting the symmetric day/night transport found by \cite{2023EGUGA..2515501B}. In our model, most of the surface in the region through which the meridional plane passes through the substellar point is land-free, with the exception of the Chukchi Peninsula in the Russian Far East (see Figure \ref{fig:map}), which disrupts the near-surface flow in the Northern Hemisphere, resulting in an asymmetric meridional overturning circulation on the ‘day side’. The cells associated with this asymmetry in the Northern Hemisphere, on both the ‘day side’ and the ‘night side’, prevent O$_3$ from accumulating near the North Pole.}

Considering Figure \ref{fig:Ox production} and Figure \ref{fig:Helmholtz_r} a) and b), we find that the O$_3$ formed at the substellar point between pressures 200 and 10 hPa (between $\sim$10 and $\sim$30 km) is first pushed towards the 120{\textdegree} longitude, then towards the poles, and from the day side to the night side by the rotationally driven winds. {At pressures $>$ 200 hPa} (below $\sim$30 km) on the night side (300° – 360° longitude), the flow from the North Pole to the South Pole (see Figure \ref{fig:Helmholtz_r} c) forces O$_3$ towards the South Pole which then gets well mixed by the rotational component of the horizontal winds resulting in high O$_3$ concentrations in the lower atmosphere near the South Pole on both the day side and the night side.

\subsection{Observational Implications}

{Our results suggest that, for an Earth-like atmospheric composition,} TRAPPIST-1e has the potential to exhibit large O$_3$ concentrations, which could be potentially observable. Furthermore, if O$_3$ is detected, this would be indicative of atmospheric oxygen. \citet{Chen_2019} demonstrated through simulated transmission spectra of M dwarf planets that JWST could potentially detect O$_3$ features during primary transit. However, detecting the prominent O$_3$ features would require over 100 transits in conditions of zero cloud coverage, and twice that number in instances of 100\% cloud coverage \citep{Lustig_Yaeger_2019}. There are numerous factors which influence the O$_3$ spectral signature, such as the line of sight of the telescope and the presence of other chemical species \citep{proedrou2016characterising}.
The brightness of an exoplanet changes as it orbits its host star and is depicted by the full-phase light curve. This curve shows the entire range of phases from minimum to maximum illumination \citep{2010exop.book.....S}. The presence of land and ocean masses can lead to non-uniform O$_3$ distributions as found in our model and hence as the planet orbits its host star, the amounts of O$_3$ that is visible might also change. Hence, the detection of O$_3$ might depend on the phase of the exoplanet we observe.
The presence of other chemical constituents in the atmosphere could also hinder O$_3$ detection by affecting its spectral signature. Chemical constituents which have similar spectral features to O$_3$ could overlap with an O$_3$ spectral features making it difficult to detect O$_3$. Finally, TRAPPIST-1 is a highly active star with dark spots and bright faculae on its surface which could make O$_3$ detection difficult by interfering with the transmission spectrum of the planet \citep{2023ApJ...955L..22L}.\\

\subsection{Limitations and Future Work}

{In this work, we have considered an Earth-like landmass distribution, orography, and atmospheric composition in order to explore the effects of landmasses on the atmospheric O$_3$ distribution, comparing our results to previous studies with slab oceans \citep{2023EGUGA..2515501B}. In the future, we would like to see studies which expand on our work, whilst also addressing the Earth-like nature of our assumptions. Suggestions for future work include:}

\begin{enumerate}

    \item{Our model assumes an Earth-like (N$_2$-O$_2$) atmospheric composition, which is a plausible but uncertain representation of TRAPPIST-1e. The actual atmospheric composition of TRAPPIST-1e remains unknown, and future observations from JWST or other missions could reveal significant differences. For now, our results provide a starting point for exploring potential atmospheric chemistry and dynamics based on an Earth-like scenario. The presence of O$_2$ is expected to lead to O$_3$ formation due to UV irradiation, which is a common feature of potentially habitable atmospheres \citep{segura2003ozone}.}

    \item {In our model, 68\% of the landmass is situated in the Northern Hemisphere. Investigating alternative land-ocean distributions - such as an even split, a majority of landmass in the Southern Hemisphere, or land configurations from different geological periods of Earth - could yield valuable insights into how land distribution impacts atmospheric circulation and O$_3$ accumulation. For example, variations in landmass distribution might influence the poleward transport of O$_3$ and subsequently affect regional atmospheric dynamics.}

    \item {In our current model, the substellar point is positioned over an ocean, leading to a relatively cold climate on TRAPPIST-1e with limited sources of liquid water. Future investigations should explore the effects of varying land-ocean distribution at the substellar point. This shift in land-ocean configuration could significantly alter the liquid ocean fraction and impact surface evaporation rates. Such changes are expected to influence the atmospheric water vapor content \citep{unknown}, potentially modifying the distribution of key atmospheric constituents, including O$_2$ and O$_3$. }

    \item {In future we could take into consideration the stellar energetic particle flux, resulting NO$_x$ production and impacts on ozone. This could potentially lead to a reduction of ozone at the poles on the night side. \citet{Chen_2020} has explored how the atmospheric chemistry on rocky exoplanets is influenced by flares.}

\end{enumerate}

\section{Summary} \label{sec:conclusions}

In this study, we used CESM2-WACCM6 GCM to simulate the atmosphere of TRAPPIST-1e, a tidally locked exoplanet, with Earth-like orography and a pre-industrial atmospheric composition, in order to {investigate the O$_3$ distribution and how it is shaped by atmospheric circulations.} Our results revealed a significant North-South asymmetry in the O$_3$ distribution, with concentrations notably higher near the South Pole at pressures $>$10 hPa (below $\sim$30 km). The day side O$_x$ production responsible for O$_3$ generation is symmetric at lower pressures (higher altitudes), but minimal at higher pressures (lower altitudes), particularly at high latitudes where the insolation is limited. Despite this, we observed a concentrated O$_3$ abundance near the South Pole. This suggested that additional factors, and not just the production of O$_x$, are shaping the O$_3$ distribution.  One factor we considered, the catalytic destruction of O$_3$ by species such as OH, HO$_2$, NO, NO$_2$, Br, and Cl was ruled out due to both the relatively symmetric distribution of said species as well as the low destruction rates. 

As such, we turned to the effects of atmospheric circulations on the O$_3$ distribution. {Our analysis revealed an asymmetry in the meridional overturning circulation.} This asymmetry is driven by the inclusion of an Earth-like landmass distribution in our model, with land-ocean boundaries shaping near surface winds to drive a flow from the North Pole to the South Pole on the night side. On the day side, the meridional overturning circulation transports O$_3$ generated at lower pressures (higher altitudes) towards the poles, whilst the rotational component of the horizontal winds carries O$_3$ from the day side to the night side. On the night side, the meridional overturning circulation transports O$_3$ from lower pressures (higher altitudes) to near the surface, where the aforementioned North-to-South flow carries O$_3$ towards the South Pole. Here, it is then mixed zonally by a wind gyre or vortex. Overall, we find TOC densities up to 28 times higher than Earth at the South Pole and 7 times higher than the Earth at the North Pole. 

{Our findings highlight the influence that tidal locking and landmass distribution, including the presence of orography, can exert on the atmospheric circulations and chemical processes of exoplanets, emphasizing the importance of considering these factors, especially land and orography, in future research.}


\begin{acknowledgments}
\nolinenumbers
We would like to thank the anonymous reviewer for their thorough review and valuable comments which helped us to improve the manuscript.
F.S.-M. would like to thank UK Research and Innovation for support under grant number MR/T040726/1. G.J.C. acknowledges the studentship funded by the Science and Technology Facilities Council of the United Kingdom
(STFC; grant number ST/T506230/1). This work was undertaken on ARC4, part of the High Performance Computing facilities at the University of Leeds, UK.
\end{acknowledgments}

\bibliography{sample631}{}
\bibliographystyle{aasjournal}



\end{document}